\documentstyle[prb,aps,multicol]{revtex} 
\begin{document}
\draft

\title{Universality of Glass Scaling in a YBa$_2$Cu$_3$O$_{7-\delta}$ 
Thin Film}
\author{Katerina Moloni, Mark Friesen, Shi Li, Victor Souw, 
P.~Metcalf, and M.~McElfresh}
\address{Physics Department, Purdue University, West Lafayette,
IN 47907-1396}
\date{\today, To be published, Phys.~Rev.~B, Dec.~1997.}
\maketitle
\begin{abstract}
The universal behavior of the continuous superconducting (glass) 
phase transition is studied in a YBa$_2$Cu$_3$O$_{7-\delta}$ thin 
film.  A new analysis technique for extracting critical exponents
is developed, using current-voltage ($J$-$E$) measurements. 
The method places narrow limits on 
the glass scaling exponents, $z_g$ and $\nu_g$, as well as the
glass transition temperature, $T_g(H)$.
The universal values $z_g = 3.8\pm 0.2$ and $\nu_g = 1.8\pm 0.1$ are obtained.  
Using these values, the data are scaled for fields in the range 1-9~T.  
Collapse of the scaling forms at different fields describes the universal 
scaling function.  The field dependence of the universal collapse is 
interpreted in terms of the 3D~$XY$ multicritical scaling theory.
\end{abstract}
\begin{multicols}{2}

\section{Introduction}
The nature of the superconducting phase transition at finite fields in high 
temperature superconductors [HTS]  remains surprisingly elusive.  In 
clean systems, recent evidence suggests a first order melting transition of 
the vortex lattice.\cite{Safar}  However in disordered 
systems, the issue is not completely resolved:  Does superconducting 
order exist at finite temperatures, separated from the normal state by a 
well-defined continuous phase transition?  Or do manifestations of 
superconductivity at nonzero temperatures represent only nonequilibrium 
phenomena, as in the flux creep description?\cite{Anderson}  
This question is of great practical interest since the two mutually exclusive 
descriptions involve distinct dynamic equations which govern the response of 
driven vortices.  In one case, the equations are 
mean-field-like;\cite{note} in the other, 
they describe a (glassy) critical state, which can occupy a very wide region 
of the superconducting phase diagram.\cite{Robustpaper}  
From a fundamental viewpoint, the debate 
encompasses the nature of the superconducting state and long range order 
in the cuprate superconductors.

Shortly after the discovery of HTS, significant progress was made to 
clarify these points.  Several types of glassy 
states\cite{FFH,BG,Bragg} were proposed 
as alternatives to the creep picture.  
A key feature of these glass phases is the true absence of resistance at low 
currents, for temperatures below the glass transition line, $T_g(H)$. 
Initial experiments strongly supported the presence of such a continuous 
phase transition,\cite{Koch,Gammel} and the superiority of the 
glass description over the creep picture.\cite{Kochcomment}  
However, further progress is hampered by the absence of a firm 
description of the glassy state, particularly regarding the nature of the 
glass transition.  Neither the critical scaling exponents nor the 
scaling functions 
have been determined analytically, due to the difficulty of the theoretical 
problem.  Further experimental treatments, while {\it individually} 
supporting a scaling description\cite{FFH} of current-voltage ($J$-$E$) 
characteristics, appear as a {\it group} 
to violate the notion of critical universality:  both sample- 
and field-dependence in the scaling exponents are observed by
some authors.\cite{seeTate}  A more rigorous test of 
universality, involving the scaling functions (which must also be universal 
according to the critical theory), has been investigated previously  
by Xenikos {\it et al.}\cite{Xenikos} 
in terms of sample-dependence, by W\"{o}ltgens
{\it et al.}\cite{Woltgens} in terms of the angle dependence of $\bf H$, 
and by K\"{o}tzler {\it et al.}\cite{kotzler} in terms of magnetic field
dependence.
Since further theoretical advancements are not 
immediately forthcoming, confirmation of the existence of true finite 
temperature superconductivity now rests on the ability to discern universal 
behavior near the purported glass transition.  
We note, however, that this test provides 
evidence only for critical scaling, 
and gives limited insight into the nature of the glassy state.

Other types of critical fluctuations (besides glass fluctuations)
may arise and compete with the 
glass excitations for critical dominance.  If the 
scaling regions overlap, the
corresponding scaling theories must be self-consistent, 
a fact which has a profound 
implications for universality.  One such competing excitation in HTS is 
associated with the 3D~$XY$ model.  The critical behavior of the zero
field superconducting transition, $T_c$, is expected to belong to the
3D~$XY$ model class, in the strongly type-II limit.\cite{FFH}  
Furthermore, the finite field superconducting transition line
$T_g(H)$ terminates at the 3D~$XY$ multicritical point:
$T_c = T_g(H=0)$.  The topology of the $H$-$T$ phase diagram 
therefore ensures that the glass and 3D~$XY$ critical regimes will overlap, 
and that their universal behaviors will be intimately related in some 
region about the 3D~$XY$ transition point.\cite{Moloni,new}  
The cautions applied to any 
mean-field dynamic description in the glass scaling region also apply to 
the 3D~$XY$ region, 
which extends over a wide temperature range at low 
fields,\cite{Salamon,Kamal}
encompassing the low field glassy region.\cite{Moloni,Friesen}

In this work, a single YBa$_2$Cu$_3$O$_{7-\delta}$ thin film is studied.  
In order to explore universal behavior, a rigorous scaling treatment is 
required to place narrow limits on the glass scaling parameters.  After 
restricting the analysis in this way, a $J$-$E$ scaling procedure is 
performed, adjusting the scaling parameters within their bounds to obtain 
the best results.  To check universality, glass scaling is compared for 
different fields, not only in terms of scaling exponents, but also scaling 
functions; both are found to be field independent.  However, fields do 
enter into the glass scaling in a way which is determined by the 3D~$XY$
 multicritical theory at low fields.  Although the present treatment 
is not applied to different samples here, it is hoped that such 
an extension may be performed in the near future.  The following 
analysis is presented in the vortex glass formulation,\cite{FFH}  
however this does not confine the critical analysis in any way,
since the different $J$-$E$ scaling formulations are isomorphic to 
one another.

\section{Experiment}
A 4000 \AA ~thick, deoxygenated YBa$_2$Cu$_3$O$_{7-\delta}$ film, 
with the $c$-axis oriented perpendicular to the film surface, was used in 
this study.  The film was prepared by pulsed laser deposition onto a 
heated (001) LaAlO$_3$ substrate, and initially had a superconducting 
transition temperature of $T_c = 92\pm 2$~K at zero applied field.  It was 
then deoxygenated at 400~$^\circ$C using a controlled flow of oxygen 
and argon gasses, followed by a rapid quench to room temperature.  
The resulting superconductor had a transition temperature $T_c = 77$~K, 
corresponding to an oxygen stoichiometry of $\delta \simeq$~0.24.  The 
film was patterned into a 120~$\mu$m $\times$ 3100~$\mu$m bridge, 
using laser ablation.

The resistivity, $\rho(T)$, was measured as a function of temperature 
for fixed currents and fields.  Measurements were then performed
for a range of applied current values 
using a conventional four point geometry, for magnetic 
fields in the range $1\leq H\leq 9~$~T.  The current densities ranged from 
20 to 4 $\times$ $10^4$~A~cm$^{-2}$.  The temperature range over which 
the resistivity 
was measured depended on the current.  Typically, data were 
collected on warming for temperature windows of 25~K, starting at temperatures 
where the voltage signal increased over the instrumental noise level.  
This ensured that 
data were collected on both sides of the glass transition temperature, $T_g$.
Throughout this work, errors are quoted as 2$\sigma$ (standard deviation) 
unless otherwise noted.  

\section{Scaling Analysis}
\subsection{Determination of $z_g$ and $T_g$}
The main focus of this work is the universal critical behavior near the 
glass transition temperature, $T_g$.  Therefore a systematic, precise 
method for determining $T_g$, and the glass scaling exponents $z_g$ 
and $\nu_g$ is required.  One method for identifying these parameters
 involves scaling of current-voltage ($J$-$E$) isotherms, obtained in 
the vicinity of $T_g$ for a single applied magnetic field ($H$).
\cite{FFH,Koch}  Typically,\cite{seeTate} the three scaling parameters
are varied simultaneously to obtain the best collapse of the isotherms.  
With no constraints placed on the fitting parameters, this method can 
obtain a reasonable scaling collapse for a large range of exponents.  
Conventionally, constraints are placed on the fitting procedure by 
identifying $z_g$ and $T_g$ from the critical isotherm,\cite{Koch} 
then requiring 
that the final scaling adjustments do not stray far from this estimate.  
However, low densities of isotherms, difficulties in 
subtracting the normal state background 
(causing the critical isotherm to appear as an imperfect 
power law), and Joule heating,\cite{Koch}
all create ambiguities in the identification of the critical isotherm, 
reducing the precision of $T_g$ to 1~K
or worse.

To improve this situation, an alternative scaling procedure is 
developed below, in which $z_g$ 
and $\nu_g$ are systematically obtained.  This procedure places 
more rigorous constraints on the fitting parameters by largely 
circumventing the difficulties described above.  An important feature 
of the method is its optimization of scaling in the vicinity of $T = T_g$; 
although this condition is required theoretically, it is not rigorously 
enforced in most scaling techniques.  A similar technique was applied 
previously in the study of 3D~$XY$, low field fluctuations in high-$T_c$ 
superconducting films.\cite{Moloni}  In the present case, the starting 
point for the scaling procedure is the following ansatz:\cite{FFH}      
\begin{equation}
\sigma (J/T)^{(z_g-1)/2} =F
\left[  t(J/T)^{-1/2\nu_g} \right] .\label{eq:a}
\end{equation}
Here, $\sigma= 1/\rho = J/E$ is the nonlinear conductivity, $F[x]$ 
is a scaling 
function, and $t = (T-T_g)/T_c$  is the relative temperature variable,
for a fixed field $H$.

Since the left hand side of Eq.~(\ref{eq:a}) displays 
no singularities, the function 
$F[x]$ must be smooth and single-valued.  In particular, it is analytic at the 
glass transition $T = T_g$, where its argument $x=t(J/T)^{-1/2\nu_g}$
(the vortex glass scaling variable) is equal to 0.  When $t = 0$, the 
scaling variable no longer depends on the current density $J$; both sides 
of Eq.~(\ref{eq:a}) should then be independent of $J$, provided that $z_g$ is 
chosen correctly.  The resistivity curves corresponding to different 
current values therefore all cross at $T_g$ when plotted as in Fig.~1.  
In the inset, the exponents $z_g$, which provide the best simultaneous 
crossing, are shown  for each applied field.  The error bars reflect the 
fact that satisfactory crossings are obtained for some range of exponents.  
Since the field dependence of $z_g$ is very weak, and since the error 
bars of $z_g$ overlap over some range 
for all the fields studied here, the average value 
$z_g = 3.8\pm 0.2$ is accepted as the result of this analysis.  Note that 
for a given $z_g$, the corresponding value of $T_g$ can be read off 
immediately from Fig.~1, with the small error of $\pm 0.02$~K.  However the 
total error in $T_g$, corresponding to the ambiguity in defining $z_g$, 
is found to be 0.4~K for the fields used here.

This {\it crossing point method} for identifying $z_g$ and $T_g$
optimizes scaling 
near $T = T_g$, and involves only the data taken near $T_g$.  In the 
analysis developed above, 
the effects of the normal state conductivity have been 
ignored, assuming the fluctuation contributions to be much larger in 
comparison.  To justify this assertion, the total non-linear conductivity 
is compared to the normal state contribution at $T = T_g$, 
which is estimated roughly by extrapolating high temperature 
conductivity data.  For the current densities used here, the total 
conductivity is at least 3 orders of magnitude larger than the normal state 
conductivity.  

\subsection{Determination of $\nu_g$}
Using the results of the previous section for $T_g$ and $z_g$, 
the static scaling exponent $\nu_g$ can now be obtained through a full 
scaling of the conductivity data according to Eq. (\ref{eq:a}).  However, it is 
beneficial to form an extension of the crossing point method of the 
previous section, which allows a value of $\nu_g$ to be identified for 
a specific (correlated)
pair of parameters ($z_g$, $T_g$).  This technique has the 
advantage of once again optimizing scaling near $T_g$.  To augment 
the crossing point method, a second (and more conventional) method 
is also used to determine $\nu_g$. This second technique complements 
the crossing point method by optimizing scaling in the ohmic 
(high-temperature) part of the scaling region.  Of the two different 
methods, the first has greater theoretical justification, since it relies only 
on data in the vicinity of $T_g$.  However, the two methods yield 
nearly identical results, with similar error levels.  This supports the fact 
that the glass critical regime is large and robust.\cite{Robustpaper}  
We now elaborate on the two methods.

{\bf Crossing point method, using data near $T = T_g$}:  The 
method of the previous section can be extended to obtain terms 
involving $\nu_g$.  Since the conductivity scaling function $F[x]$ of 
Eq.~(\ref{eq:a}) is analytic at $x = 0$, a Taylor series expansion may be 
performed about this point.  For $t \simeq 0$, Eq.~(\ref{eq:a}) becomes
\begin{eqnarray}
\sigma (J/T)^{(z_g-1)/2} & = & F[0]+t(T_g/J)^{1/2\nu_g}
F' [0] +\cdots  \nonumber \\
& = & F[0]+mt+\cdots . \label{eq:b}
\end{eqnarray}
Truncating Eq.~(\ref{eq:b}) at its zeroth order term, $F[0]$, gives the $z_g$ 
crossing point method used above.  To determine $\nu_g$, the 
first order term is also retained.  The coefficient 
$m=(T_g/J)^{1/2\nu_g}F' [0]$
of the linear term in Eq.~(\ref{eq:b}) 
is determined for different current densities 
$J$ at  $T = T_g$ by observing the slopes of the different curves in Fig.~1 
at the crossing point.  The result found above, $z_g= 3.8$, is used on the 
left hand side of Eq.~(\ref{eq:b}).  Since $\nu_g$ should be independent of the 
current, the values obtained in this way are averaged over the different 
currents to obtain error bars.  The results of the crossing point method 
for $\nu_g$ are shown in Fig.~1 (inset) for each field.  
No obvious trend of $\nu_g$ 
with the field is apparent in this figure (and none is expected).  It is 
therefore meaningful to average $\nu_g$ over field variations, giving 
a final result of $\nu_g = 1.77\pm 0.03$.  Keeping significant figures 
consistent with $z_g$, we quote $\nu_g = 1.8\pm 0.1$.          

{\bf Slope method, using data in the ohmic regime}:  This makes use of 
the ohmic resistivity data in the limit of $J\rightarrow 0$ and $T > T_g$.  
Typically, data are obtained here using current densities in the range 
20 to  $5 \times 10^2$~A~cm$^{-2}$.  In the ohmic regime, $\rho$ is 
independent 
of $J$, so Eq.~(\ref{eq:a}) reduces to $\rho \sim t^{\nu_g(z_g-1)}$.  
This power law behavior 
can be observed in Fig.~2.  The slope of the linear (power law) portion 
of the curve can then be used to determine $\nu_g(z_g-1)$, and 
subsequently $\nu_g$, using $z_g = 3.8$.  The apparent power law 
behavior of $\rho$, observed over two decades in Fig.~2, indicates 
that data used to obtain $\nu_g$ in this way remain within the critical 
scaling region. The resulting $\nu_g$ values are shown for different 
fields in the inset of Fig.~2.  The error bars for each data point reflect 
the ambiguity in identifying the limits of the ohmic scaling region.  
Similar to the inset of Fig.~1, no obvious trend of $\nu_g$ with $H$ is 
observed, nor is there any correlation between the field variations in 
the insets of Figs.~1 and~2.  
This provides additional support for identifying $\nu_g$ 
as a field average: $\nu_g = 1.79\pm 0.04$, a result nearly identical to 
the crossing point method.  Again, keeping significant figures 
consistent with $z_g$, we quote $\nu_g = 1.8\pm 0.1$.  This value is 
used in the following analysis.

\section{Discussion of Universality}
The absence of field dependence in the exponents $\nu_g$ and $z_g$, 
demonstrated in Section~III, is consistent with the vortex glass theory, 
which states that the critical behavior of the linear or nonlinear 
conductivity $\sigma$ near the transition line $T_g(H)$ should belong to 
a single universality class, for some range of $H > 0$.\cite{FFH}  
The vortex glass excitations are thought to be formed within a medium 
of field-induced vortices, whose excursions from perfect rigid rod 
geometry, parallel to ${\bf H}$, are affected by entropy, applied currents, 
and the pinning effects of quenched disorder.  The density of field 
vortices varies with $H$, while quenched disorder does not.  For a 
particular sample then, vortex matter at different fixed fields forms 
unique vortex systems.  The behavior and response of these unique 
systems to external probes may vary greatly with the field.  Indeed, 
it has been noted that changes in the field may be responsible for 
phenomena such as dimensional crossover, causing vortex matter 
at low and high fields to behave very differently.\cite{BFGLV}  
However, many of these phenomena are of mean-field origin, and  
should not be observed in measurements which predominantly reflect a
critically diverging correlation length.
It is not surprising then that vortex glass universality, observed 
here for all studied fields, $H > 0$, along $T_g(H)$ (and in 
Ref.~\onlinecite{Moloni} up 
to much higher fields) appears to hold in spite of such variations
with field.

In order to observe universal glass behavior of both the critical exponents 
and the scaling function, it is necessary to introduce a new notation to 
account for field dependence in the vortex glass description.  This field 
dependence is {\it not} specified by the vortex glass theory; it is not 
known to be universal, {\it a priori}, 
and may vary from sample to sample.  The 
conductivity scaling of Eq.~(\ref{eq:a}) can then be rewritten as
\begin{equation}
\left( \frac{\rho}{\rho_0} \right) 
|t|^{-\nu_g(z_g-1)} = G_\pm
\left[ \frac{J}{T} \frac{T_g}{J_0} |t|^{-2\nu_g} \right]
. \label{eq:c}
\end{equation}
Here, we have introduced the field dependent resistivity scale 
$\rho_0(H)$ and the current scale $J_0(H)$.  The transition line 
$T_g(H)$ is also field dependent. The new functions $G_{\pm}[x]$,
corresponding to $T>T_g$ and $T<T_g$, 
and their arguments, are dimensionless.

To further develop a universal description, it is necessary to identify 
the scales $\rho_0(H)$ and $J_0(H)$ empirically.  
Methods for obtaining characteristic 
resistivity and current scales exist in the literature,\cite{Koch} 
however we believe the following definitions to be the most objective: 
\begin{eqnarray}
\rho_0 & =  & \left. \rho t^{-\nu_g (z_g-1)} 
\right|_{T>T_g,J\rightarrow 0} ,\label{eq:d} \\
J_0 & = & \left. J (\rho /\rho_0)^{2/(1-z_g)}
\right|_{T=T_g} ,\label{eq:e}
\end{eqnarray}
where $\rho$ and $J$ refer here to experimentally determined quantities.
In this definition, $\rho_0$ is the ohmic resistivity scale, while $J_0$ 
is the current required to produce a resistance $\rho = \rho_0$ along the 
critical isotherm.

Using these definitions, the nonlinear resistivity curves, scaled at 
different fields according to Eq.~(\ref{eq:c}), 
collapse as shown in Fig.~3.  
The choice of normalization in
Eqs.~(\ref{eq:d}) and (\ref{eq:e}) causes the dimensionless 
point $(1,1)$ to be located as shown in the figure.  We emphasize that this 
collapse of $J$-$E$ scaling curves, including different fields, 
demonstrates the universality of the
vortex glass transition.  By varying $\nu_g$, $z_g$, and 
$T_g(H)$ within their error bars, as found above, it is not possible to 
obtain any discernible improvement  in the scaling beyond Fig.~3.
Additionally, for $z_g<3.8$, the scaling deteriorates markedly.

The obtained field dependences of $\rho_0$ and $J_0$
are shown in Fig.~4.  As mentioned above, neither of these 
quantities is specified by the vortex glass theory.  However, they can be 
determined in a multicritical theory such as 3D~$XY$, which is thought 
to apply to topological fluctuations at low fields.\cite{FFH,Moloni,Salamon}  
The following power law dependencies of $\rho_0(H)$ and $J_0(H)$ are 
derived from the 3D~$XY$ theory:\cite{Friesen}
\begin{eqnarray}
\rho_0 (H) & = & \bar{\rho}
H^{[\nu_{xy}(z_{xy}-1)-\nu_g(z_g-1)]/2\nu_{xy}} ,\label{eq:f} \\
J_0 (H) & = & \bar{J} H^{1-(\nu_g/\nu_{xy})} ,\label{eq:g}
\end{eqnarray}
where $\nu_{xy}$ and $z_{xy}$ are 3D~$XY$ scaling exponents, 
and $\bar{\rho}$ and $\bar{J}$ are unknown constants which are sample-, but not 
field-dependent.  

In Fig.~4, $\rho_0(H)$ and $J_0(H)$ are described well by power laws for 
low fields.\cite{note10}  It is important to note that,
in contrast to other 3D~$XY$ analyses performed in the literature, 
the observance of this 
power law does not hinge on an appropriate choice of $T_c$ as a fitting
parameter.  In the fits shown, only data are used 
which fall within the expected 3D~$XY$ scaling region for this 
sample\cite{Moloni} ($H<8$~T).  We find corresponding slopes of
$-3.1\pm 0.2$ ($\rho_0$) and $-1.39\pm 0.04\simeq -1.4$ ($J_0$).
A self-consistency test is provided by 
Eq.~(\ref{eq:g}), using the known value of $\nu_{xy}=0.67$.\cite{zinn}
The obtained value of $\nu_g=1.60\pm 0.03\simeq1.6$ provides a  
satisfactory check on the universality analysis.  Additionally, 
$z_{xy}$ may be calculated from Eq.~(\ref{eq:f}), using the slope
obtained in Fig.~4, together with $\nu_{xy}=0.67$, $\nu_g=1.8$, and 
$z_g=3.8$.  The result is $z_{xy}=2.3\pm 0.2$. 
[Note that the error sources in Eq.~(\ref{eq:f}) are correlated.] 
This result is closer to that given
by Booth {\it et al.}\cite{booth} than Moloni {\it et al.}\cite{Moloni}
We speculate that the difference between the present results and those of
Ref.~\onlinecite{Moloni} may reflect the difficulty in treating 
background contributions to $\sigma$, which impedes the analysis
of Ref.~\onlinecite{Moloni},
but not the present one.

\section{Conclusions}
In this work, the vortex glass scaling parameters $\nu_g = 1.8\pm 0.1$, 
$z_g = 3.8\pm 0.2$, and $T_g(H)$ have been obtained through a 
systematic scaling technique, involving the vortex glass crossing point at 
$T = T_g$.  Similar to other deoxygenated YBa$_2$Cu$_3$O$_{7-\delta}$ 
films, the vortex glass scaling region was found to be wide, on the order 
of 10~K at $H = 7$~T.  The absence of field dependence in the exponents 
$\nu_g$ and $z_g$ provides a demonstration of universal behavior, since 
different fields correspond to unique vortex systems.  A second (and stronger) 
demonstration of universality is given by the collapse of scaled $J$-$E$ 
curves obtained at different fields in the vortex glass critical region, 
as shown in Fig.~3.  Because universality is an essential component of 
critical phenomena, these checks on both the glass exponents and the scaling 
functions provide strong evidence for a critical phase transition line along 
$T_g(H)$ for $H > 0$.  While additional evidence has not been presented 
here to demonstrate that this transition involves explicitly the broken 
symmetry associated with a vortex glass, it should be noted that the scaling 
functions in Fig.~3, below and above $T_g(H)$, exhibit the expected glassy 
characteristics at low currents:  below $T_g(H)$ the non-linear resistivity 
decays exponentially to 0, while above $T_g(H)$ the resistivity decays 
exponentially to its normalized value of 1.

The values reported here for the dynamic scaling exponent, $z_g$, and 
the static scaling exponent, $\nu_g$, fall within the range of previously 
reported values,\cite{seeTate} although our
result $z_g = 3.8\pm 0.2$ falls barely 
within the expected range\cite{FFH} $z_g\geq 4$.  The significant
variation of reported $z_g$ values 
in the literature could imply that 
conventional methods for determining $z_g$ may 
involve systematic errors in the identification of the critical isotherm
(and thus $T_g$), a problem over which we believe the techniques applied
here provide greater control.
Alternatively, the field variations of $z_g$ could result from
the glass correlation length becoming smaller than the 
distance between field-induced vortices.\cite{Koch,seeTate}
Finally, $z_g$ variations could imply an absence of 
dynamic universality.  These possibilities are not investigated 
in further detail
here.  However, to study this issue in more depth, it will 
be important to apply the methods developed in this work to other 
high-$T_c$ samples.  In each case, universality must be demonstrated 
for both the scaling exponents and the functions.

Finally, through Eqs.~(\ref{eq:f}) and (\ref{eq:g}), 
evidence has been presented which, 
together with analyses such as Ref.~\onlinecite{Moloni}, provides strong 
support for 
3D~$XY$ multicritical scaling.  The present analysis involves scaling of 
the {\it nonlinear} resistivity, and therefore complements previous 
3D~$XY$ studies,\cite{Moloni,Salamon,Howson} which apply only to the 
{\it linear} resistivity.  The focus of the present work has been the vortex 
glass critical regime, for which Eqs.~(\ref{eq:c}), (\ref{eq:f}), 
and (\ref{eq:g}) augment previous 
analyses by explaining the field dependence of vortex glass scaling, in 
addition to the temperature dependence.  
Through these equations it is found that $z_{xy}\simeq 2.3$,
in agreement with Ref.~\onlinecite{booth}; the present result does
not suffer from background subtraction difficulties.  Eqs.~(\ref{eq:f}) 
and (\ref{eq:g}) apply only to 
the 3D~$XY$ scaling region, thus allowing the range of 3D~$XY$ 
scaling to be determined.  For the sample studied here, the 3D~$XY$ 
critical regime extends to 8~T along the line $T = T_g(H)$.
If the dynamics of the glass transition are ultimately determined to be
non-universal, it will be
important to ascertain whether the self-consistency of 3D~$XY$ multicritical
scaling, demonstrated here, still persists.

\section*{Acknowledgments}
We thank Paul Muzikar and S. Salem-Sugui for many helpful discussions,
and we are indebted to J. Deak for a careful reading of this manuscript.
This work was supported through the Midwest Superconductivity 
Consortium (MISCON) DOE grant No. DE-FG02-90ER45427, and the 
Materials Research Science and Engineering Center (MRSEC) Program 
of the NSF under Award No. DMR-9400415.

\begin{figure}
\end{figure}
\noindent FIG.~1. 
Crossing point method at $H = 1$~T.  Different curves correspond to 
different current densities.  The curves all cross at a single point, 
$T = 65.2$~K, provided $z_g$ is chosen correctly, thus identifying $T_g$.  
The $z_g$ values obtained in this way are shown in the inset for each 
applied field.  Additionally, by determining the slopes of the curves
at $T_g$, the exponent $\nu_g$ can be 
obtained, as described in the text.  These $\nu_g$ values are shown 
for each applied field in the inset.

\begin{figure}
\end{figure}
\noindent FIG.~2.
Slope method for $H = 1$~T.  The static scaling exponent $\nu_g$ 
is calculated from the power-law dependence of the ohmic $\rho (T)$ 
data.  Values of $T_g$ and $z_g = 3.8$ used in this fit 
were obtained from the 
crossing point method of Fig.~1.  Resulting values of $\nu_g$ are shown 
in the inset for each applied field.  

\begin{figure}
\end{figure}
\noindent FIG.~3.
Scaled $\rho (J,T)$ data from three different fields.  ($\tilde{\rho}$
and $\tilde{J}$ represent, respectively, the left hand side of 
Eq.~(3), and 
the argument ($x$) of the scaling function $G_\pm[x]$. 
Dimensionless, arbitrary units are used.) 
The quantities $\rho_0$ and $J_0$ of Eq.~(3)
have been found for each field so that the three 
curves collapse onto a single, universal curve.  The scaling exponents 
$z_g = 3.8$ and $\nu_g =1.8$ are used for all fields.  Although the 
collapse is excellent for all five fields used here, only three are shown 
for clarity. 
The dashed lines represent asymptotic behaviors, and their crossing 
identifies the normalized point $(1,1)$; see Eqs.~(4) and (5).

\begin{figure}
\end{figure}
\noindent FIG.~4.
Field dependence of the quantities $\rho_0$ and $J_0$.  The dashed 
lines represent best fits, involving data points only from the 
expected 3D~$XY$ scaling region\cite{Moloni} ($H<8$~T).

\end{multicols}

\begin{references}
\bibitem{note10}Similar power laws were observed 
	previously,\cite{kotzler} but were not attributed to
	3D~$XY$ critical phenomena.
\end{references}
\end{document}